\documentclass[12pt]{article}
\usepackage[dvips]{graphics}
\renewcommand{\L}{{\cal{L}}}
\begin{document}
\baselineskip=6.0mm
\begin{titlepage}
\begin{flushright}
  KOBE-TH-98-02\\
  hep-th/9805067
\end{flushright}
\vspace{2.3cm}
\centerline{{\large{\bf The Gauge Hierarchy Problem}}}
\centerline{{\large{\bf and}}}
\centerline{{\large{\bf Higher Dimensional Gauge Theories}}}
\par
\par
\par\bigskip
\par\bigskip
\par\bigskip
\par\bigskip
\par\bigskip
\renewcommand{\thefootnote}{\fnsymbol{footnote}}
\centerline{{\bf Hisaki 
Hatanaka}$^{(a)}$\footnote[1]{e-mail:hatanaka@oct.phys.kobe-u.ac.jp},
{\bf Takeo  Inami}$^{(b)}$\footnote[2]{e-mail:inami@ieyasu.phys.chuo-u.ac.jp}, 
and {\bf C.S. Lim}$^{(c)}$\footnote[3]{e-mail:lim@oct.phys.kobe-u.ac.jp}}
\par
\par\bigskip
\par\bigskip
\centerline{$^{(a)}$ Graduate School of Science and Technology, Kobe University, Kobe 657, 
Japan} 
\centerline{$^{(b)}$ Department of Physics, Chuo University, Bunkyo-ku, Tokyo 112, Japan}
\centerline{$^{(c)}$ Department of Physics, Kobe University, Nada, Kobe 657, 
Japan}
\par
\par\bigskip
\par\bigskip
\par\bigskip
\par\bigskip
\par\bigskip
\par\bigskip
\par\bigskip
\par\bigskip
\par\bigskip
\centerline{{\bf Abstract}}\par
We report on an attempt to solve the gauge hierarchy problem
in the framework of higher dimensional gauge theories.
Both classical Higgs mass and quadratically divergent quantum  correction to 
the mass are argued to vanish. Hence the hierarchy problem in its original sense is solved. The remaining finite mass correction is shown to depend crucially 
on the choice of boundary condition for matter fields, and a way to  fix it dynamically is presented. We also point out that on the simply-connected space $S^2$ even the finite mass correction vanishes.

\par\bigskip
\par\bigskip
\par\bigskip
\par\bigskip
\par\bigskip
\par\bigskip
\par\bigskip

\end{titlepage}
\newpage

\newcommand{\mass}{m_{H}}

\vspace{0.5 cm}
\leftline{\bf 1. Introduction}
\vspace{0.2 cm}

The Standard Model of electroweak theories has been very successful in describing 
observed phenomena. However,  a serious
problem arises when it is regarded as the
effective low energy theory of a certain more fundamental theory.  Suppose that the Standard Model is valid up to a
physical momentum cutoff $\Lambda $, where it should be replaced by a more
fundamental theory. When the cutoff is huge compared with the weak
scale, $ \Lambda \gg M_W $, the huge mass scale $\Lambda$ may
potentially disturb the low energy $(\sim M_W)$ physics. This is the
 so-called gauge hierarchy problem.  It has played crucial  role in particle
physics;  the attempts at solving the problem have led to
 theories beyond the Standard Model, such as supersymmetric models
\cite{Sakai}, models with dynamical gauge symmetry breaking\cite{techni},
and an idea that the Higgs particle should be regarded as a
pseudo-Nambu-Goldstone boson \cite{Inoue}.

There are actually two kinds of gauge hierarchy problem: (i) When the
 fundamental theory is of the type of grand unified theory, the problem
arises already at the tree level.  For instance, in the  $SU(5)$ GUT, in order
to keep the mass of doublet component of the 5-plet Higgs to the weak scale,
while that of color triplet partner to the GUT scale, one must
fine-tune the parameters in the scalar potential to the precision of
$(M_W/M_{GUT})^2 \sim 10^{-26}$: the triplet-doublet mass splitting problem. 
(ii)  Even if the mass of the Higgs doublet is tuned to be
small at the tree level, the mass-squared will suffer from huge quantum
corrections proportional to $\Lambda^2$: the problem of ``quadratic
divergence''. Again one has to fine-tune the bare mass parameter to the
precision $ \sim 10^{-26}$.  We note that the SUSY theory, an attractive candidate
for the ``new physics'', does not immediately solve the first problem, though it solves the second one successfully.

The hierarchy problem is a clue to the search for the candidates for new physics. Hence 
to exhaust possibilities of the solution to the hierarchy problem is highly  desirable.  The guiding principle in the investigation in this direction is the concept of the
``naturalness'' by 't Hooft \cite{'tHooft}; the smallness of some
physical quantity is naturally ensured provided the symmetry of the
theory is enhanced when the quantity vanishes.
In this paper we shall investigate a new possibility to solve the
hierarchy problem where the smallness of the Higgs mass is naturally
guaranteed by the local gauge symmetry of a higher dimensional gauge
theory. 

The idea is the following. Let $A_M (M=0$ to $D-1)$ be a gauge field 
in the  $D$ dimensional space-time, and decompose it into
4-dimensional and extra space components, 
\begin{equation} 
A_{\mu} (\mu=0,... ,3), \ \ \   A_m(m= 4,... ,D-1).
\end{equation}
After the compactification of
the extra space, the extra space component $A_m$ behaves as a set of
scalar fields in our 4-dimensional world, which we regard as our
Higgs fields. Apparently, the naturalness condition stated above is
satisfied, since when the mass of $A_m$ vanishes the local gauge
symmetry with respect to the extra space coordinates arises. Thus the second
problem of quadratic divergence should be solved. Moreover at the tree
level the local gauge invariance automatically forbids the presence of the
Higgs mass, and the problem (i) is also readily solved once a viable
(GUT-type) model is constructed. By taking a toy model we will confirm
this assertion, finding that a mechanism works to eliminate the
ultraviolet quadratic divergence thanks to the local gauge invariance.
 We generally get a finite mass correction, and it depends on
the global property of the compactified space.
It is widely understood that the supersymmetry plays an essential
role for the vanishing quadratic divergence \cite{SUSY}, even in higher dimensional supersymmetric 
Yang-Mills theories. In our view, it is worthwhile
clarifying the question whether the divergence can also be eliminated by the
mechanism advocated here relying on the gauge symmetry.

Another interesting aspect of studying higher dimensional gauge theories
is the possibility for the non-trivial topology of the compactified
extra space to affect physics, especially via the boundary conditions
(b.c. for short) imposed on fields along the directions of
$y_m$. Casimir effect in QED is a typical example.  Quantum corrections have been studied 
in theories with compactified space \cite{Tom},\cite{Ford},\cite{Inami}. 
In particular, it has been pointed out that in the
space-time of $M^3 \times S^1$ a photon would propagate faster
than the speed of light (!), i.e. $m_{\gamma}^2 < 0$, a  consequence
of the periodic b.c. of fermion field with respect to the direction 
$S^1$ \cite{Tom},\cite{Ford}. This result suggests the importance of quantum fluctuations in the
compactified extra space, and that the b.c. of the fluctuating fields
does affect the physics.  In fact, we will discuss that the remaining
finite quantum correction to the Higgs mass-squared crucially depends on
the choice of the b.c.. It will also be argued that, since the
compactified space $S^1$ is nonsimply-connected space, the b.c. may be
reduced to the effect of constant background of $A_m$, a sort of
Aharonov-Bohm (A-B)effect, and therefore can be fixed dynamically
\cite{Hosotani}.

A remark is in order concerning the tower of massive Kaluza-Klein modes. The
invariance under local gauge transformations with gauge parameters
depending on the coordinates $y^m$ is essential for our mechanism to 
work. To keep
this gauge symmetry, the tower of massive modes have to be 
taken into account. This is because, after a gauge transformation, e.g.  of
 a fermion field,

\begin{equation}
\psi  \rightarrow \psi^{\prime}= e^{i\epsilon^{a}T^{a}} \psi,
\end{equation}
with a gauge parameter $\epsilon^{a}(y^m)$
depending only on the extra space coordinate $y_m$, $\psi^{\prime}$
 necessarily depends on $y^m$ and hence contains higher modes (Fourier modes in 
the case of $S^1$) in $y^m$. 
Our proposal of considering the local gauge symmetry with respect to 
the coordinate $y^m$ differs from the usual view in which only 
massless modes (zero modes) are assumed to take part in the compactified 
theory.  
 
By local gauge transformations in compactified gauge theories one usually 
refers to local gauge transformation with respect to the 4-dimensional 
space-time coordinates $x^{\mu}$. In the present work local gauge transformations mean gauge transformations which depend on the higher dimensional coordinates $x^{\mu}, y^m$. We focus more specifically 
on the gauge transformations Eq.(2) which depend on the coordinate 
$y^m$.  

\vspace{0.5 cm}
\leftline{\bf 2. A toy model and the quantum correction to the Higgs mass}
\vspace{0.2 cm}

We begin by showing the mechanism for the disappearance of the
quadratically divergent quantum correction to the Higgs mass  in a toy model: $D+1$ dimensional QED in the space-time
$M^D \times S^1$, the product of D-dimensional Minkowski space-time and a
 circle with a radius $R$, whose coordinates are $x^{\mu}$
and $y$, respectively. We present the computation for an arbitrary 
dimension $D$, and the case of our main interest, $M^4 \times S^1$, is 
obtained by setting  $D = 4$.  The field contents are the $D+1$ dimensional photon $A_M$ and an electron
$\psi$ with mass $m$.  The gauge field $A_M$
decomposes into D-dimensional and extra-space components, $A_M = (A_{\mu},A_y)$. 
Only zero modes are considered as external (real) states, while 
massive modes have to be taken account of in the intermediate states. 
The latter will play essential role in the quantum corrections. 
 
At the tree level the Higgs mass
$\mass$ vanishes, i.e. $\mass = m_{A_y} = 0$, due to the local gauge invariance.  Thus the first hierarchy
problem is readily solved.  Our next task is to calculate the quantum
correction to $\mass^2$, to see whether the quadratic divergence 
disappears, as we naively expect from the fact that the photon mass
remains zero under quantum corrections in ordinary QED, again due to
the gauge invariance. 

We begin by recapitulating the $\mass^2$ at the lowest order in
the limit of $R \rightarrow \infty$ , i.e., in $M^{D+1}$ space-time.
As is well-known,  $\mass^2$ vanishes in this limit:

\begin{eqnarray}
\mass^2
&=&  \left(\frac{2^{\frac{D+1}{2}}}{D+1}\right)ie^{2}L \int 
\frac{d^{D+1}k}{(2\pi)^{D+1}}
 \left\{
 (1-D)\frac{1}{k^2 - m^2} + 2m^2 \frac{1}{(k^2-m^2)^2}
 \right\} \nonumber \\
&=& \left(\frac{2^{\frac{D+1}{2}}}{D+1}\right)\frac{e^{2}L}{(4\pi)^{\frac{D+1}{2}}}
 \left\{
 (1-D) + 2m^2 \frac{\partial}{\partial m^2}
 \right\}
 \Gamma \left(\frac{1}{2} - \frac{D}{2}\right)(m^2)^{\frac{D-1}{2}} 
\nonumber \\
&=& 0 , 
\end{eqnarray}
where $L \equiv 2\pi R$ and the D-dimensional charge $e$ is related to the original charge 
$e^{(D+1)}$ as $e = L^{-1/2}e^{(D+1)}$. 

Next let us calculate $\mass^2$ in the manifold of our interest, $M^D
\times S^1$ with a finite radius $R$ of the circle.  The
 requirement that physical quantities should be single-valued functions of
space-time coordinates demands that the gauge field $A_M$ should be
single-valued. However, the electron may have an arbitrary b.c., as
only the product $\bar{\psi}\psi$ is of physical relevance:
\begin{eqnarray}
\psi(x_\mu, y + 2\pi R) = e^{i\alpha} \psi(x_\mu, y),
\end{eqnarray}
where $\alpha$ is an arbitrary phase.
The Higgs mass-squared $\mass^2$ calculated for the b.c. is given as
\begin{eqnarray}
\mass^2 = ie^2 2^{\frac{D+1}{2}} \int 
\frac{d^Dk}{(2\pi)^D} \sum_{n}
\left\{
-\frac{1}{(\frac{2\pi n + \alpha}{L})^2 + \rho^2}
+ 2\rho^2 \frac{1}{[(\frac{2\pi n +\alpha}{L})^2 + \rho^2]^2}
\right\},
\end{eqnarray}
where $\rho^2 \equiv -k^\mu k_\mu + m^2$. 
It should be noted that the discrete
momentum in y-direction gets a constant shift proportional to $\alpha$
due to the arbitrarily chosen non-trivial b.c. Eq.(4): $k_y = \frac{2\pi
n + \alpha}{L}$ (n : integer). 

The above expression for $\mass^2$ is
superficially highly divergent. However, a finite expression is obtained by 
subtracting zero from
Eq.(5), i.e.  by subtracting the contribution for $L \rightarrow \infty$,
Eq.(3).  By utilizing a relation
\begin{eqnarray}
\sum_{n} \frac{1}{(\frac{2\pi n + \alpha}{L})^2 + \rho^2}
-
L \int \frac{dk_y}{2\pi} \frac{1}{k_{y}^{2} + \rho^2}
=
\left( \frac{L}{2\rho} \right)
\left[
\frac{\sinh (\rho L)}{\cosh(\rho L)- cos\alpha} -1
\right] ,
\end{eqnarray}
we arrive at (after Wick rotation) 
\begin{eqnarray}
\mass^2
&=&
-ie^{2}2^{\frac{D+1}{2}} \int \frac{d^Dk}{(2\pi)^D}
\left( 1+ \rho \frac{\partial}{\partial\rho}\right)
\left( \frac{L}{2\rho} \right)
\left[
\frac{\sinh(\rho L)}{\cosh(\rho L) - \cos\alpha} -1
\right] \nonumber \\
&=&
\frac{1}{2^{\frac{D-1}{2}} \pi^{\frac{D}{2}} \Gamma(\frac{D}{2})}
e^{2}L^{2} \int_0^\infty dk \ k^{D-1}
\frac{1-\cosh(\sqrt{k^2 + m^2}L) \cos\alpha}{[\cosh(\sqrt{k^2 + m^2}L)-\cos\alpha]^2} \nonumber \\ 
&=&
\frac{1}{2^{\frac{D-1}{2}} \pi^{\frac{D}{2}} \Gamma(\frac{D}{2})}
e^{2}L^{-D+2} \int_0^\infty ds \ s^{D-1}
\frac{1-\cosh\sqrt{s^2 + (mL)^2} \cos\alpha}{[\cosh\sqrt{s^2 + (mL)^2}-\cos\alpha]^2} ,
\end{eqnarray}
where a dimensionless variable $s = kL$ has been introduced. As expected,
no divergence appears in the quantum correction to $\mass^2$, for
arbitrary dimension D; the integral is super-convergent, i.e. for larger
$k$ the integrand behaves as $-2(\cos\alpha)k^{D-1} e^{-\sqrt{k^2 + m^2}L}$ , where $e^{-\sqrt{k^2 + m^2}L}$ is analogous to the Boltzman factor
in the case of finite temperature QED where $\psi$ has an anti-periodic
b.c. ($\alpha = \pi$) in the time direction.
Let us note that if we take only the zero mode $n = 0$ into account in
Eq.(5), the calculation reduces to that of the pseudo-scalar 2-point
function in the ordinary 4-dimension (for $D = 4$) and therefore
the result diverges quadratically.

We thus have solved the second hierarchy problem of quadratic
divergence, and the gauge hierarchy problem in its original sense has been solved. 
 We note, however, there remains a finite
correction to $\mass^2$.  In the usual argument the gauge hierarchy problem is concerned with U.V. behavior of the theory. By contrast, the appearance of finite $\mass^2$ depends on the global nature of the 
extra space, i.e., whether it is infinite or compact, and is controlled by the I.R. nature of the theory.   

When L is relatively large, such I.R. effect is expected to be small,
roughly of the order $(1/L)^2$.  Thus for $L \geq 1/(1\mbox{TeV})$
\cite{Antoniadis} the hierarchy problem is totally solved. 
It is noteworthy that such large (length) scale compactification has 
recently attracted revived interest in the context of GUT and/or 
superstring \cite{Large}. 
For much smaller L, say of ${\cal O}(1/M_{GUT})$ or ${\cal O}(1/M_{pl})$,
such ``I.R.'' effect is expected to become large, in general.
To be precise, $\mass^2$ also depends on m and the boundary condition .

In order to see how $\mass^2$ behaves as a function of L and m,
we first study a specific case of periodic boundary condition,
 i.e. $\alpha=0$,
where $\mass^2$ is simply given by
\begin{eqnarray}
\mass^2 = - \frac{1}{2^{\frac{D-1}{2}}\pi^{\frac{D}{2}}\Gamma 
(\frac{D}{2})}
e^{2}L^{-D+2} \int_0^\infty ds\quad \frac{s^{D-1}}{\cosh\sqrt{s^2 + (mL)^2}-1}.
\end{eqnarray}
%
%
\begin{figure}
\begin{center}
\includegraphics{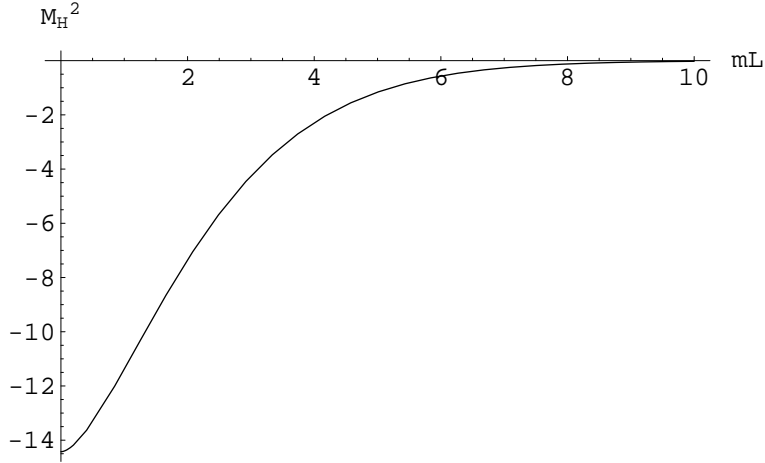}
\caption{ The $\mass^2$ measured in the unit of 
$e^{2}/(2\sqrt{2}\pi^2L^2)$ as a  function of $mL$ in the case of 
periodic b.c..}
\end{center}
\end{figure}
Fig.1 shows, for the case of $D = 4$, the behavior of $\mass^2$  as a function of $mL$(electron
mass measured in the unit of $1/L$).  As is
seen there $\mass^2$ indicates an exponential damping (for large $mL$),
essentially due to the aforementioned ``Boltzman factor''.  Thus, if $mL$
 is as large as $\sim$ 30, the gauge hierarchy is understood naturally as $M_W^2
\sim (1/L)^2 e^{-mL}$ (geometrical hierarchy). For smaller values of 
$mL$ the $\mass^2$ suffers from a large correction of ${\cal O}(1/L^2)$.

\vspace{0.5 cm}
\leftline{\bf 3. The effect of boundary condition
and its dynamical determination}
\vspace{0.2 cm}

Next we will investigate how the boundary condition
(b.c.)$\alpha$  for $\psi$ affects $\mass$. 
%
%
\begin{figure}[htbp]
\begin{center}
\includegraphics{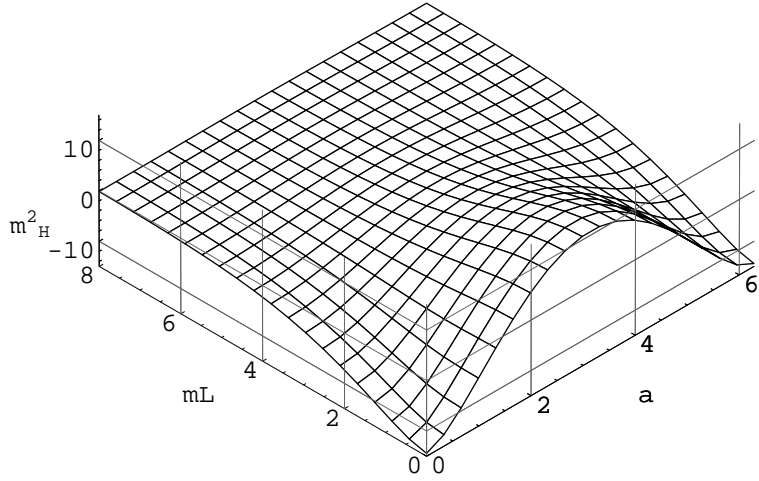}
\caption{
The 3D plot of $\mass^2$ (in  the unit of $e^{2}/(2\sqrt{2}\pi^2L^2)$)
 as a function of $a$ and $mL$, with $a$ standing for $\alpha$.}
\end{center}
\end{figure}
In Fig.2 we plot the result of Eq.(7) as a function of both $\alpha$ and $mL$.  It is interesting to note that $\mass^2$ even changes its sign as $\alpha$ varies. 
In fact, we find from Eq.(7) that $\mass^2 < 0$ for $\alpha = 0$ and  $\mass^2 > 0$ for $\alpha = \pi$ irrespective of $mL$. 
Specifically, $\mass^2 = 0$ is realized at $\alpha \simeq 0.46 \pi$ 
for the case of $m = 0$. 

So far the b.c. has been put by hand, and we feel it uncomfortable 
that the physical quantity $\mass$ depends on the b.c.. 
We will now attempt to attribute some physical meaning to the phase $\alpha$
and will study the possibility of dynamically fixing it.  In order to consider the
physical relevance of the phase $\alpha$ it is worthwhile noting
that if we wish we can always rotate away the phase $\alpha$ by re-phasing
of $\psi$, since the theory has local U(1) gauge symmetry:
\begin{eqnarray}
\psi(x_\mu, y) \rightarrow \psi'(x_\mu, y) &=& U \psi(x_\mu,y), \\
U &=& e^{-i\frac{y}{L}\alpha} \in U(1). 
\end{eqnarray}
The field $\psi'$ obeys the periodic b.c.:
\begin{eqnarray}
\psi'(x_\mu,y + L) = \psi'(x_\mu,y).
\end{eqnarray}
On the other hand, $A_y$ gets a constant shift as
\begin{eqnarray}
A_y \rightarrow A'_y
 = A_y - \left(\frac{1}{e}\right)\frac{\alpha}{L} , 
\end{eqnarray}
i.e., only the zero-mode of $A_y$ is affected.
Thus, without loss of generality, we may always assume that the field $\psi$ obeys the periodic b.c., while $A_y$ may have a constant background, or a vacuum expectation value, $\langle A_y\rangle $.   
We have already seen in the Eq.(8) that $\mass^2$ is negative for periodic b.c. \cite{Ford}. This clearly suggests that the real vacuum state is not at $\langle A_y\rangle=0$;  
we are led to consider nonvanishing $\langle A_y\rangle $.

Naively thinking, such constant shift of the higher dimensional gauge 
field has no physical effect.
In fact if $A_y=0$ and $F_{\mu\nu}=0$, corresponding to the naive vacuum 
$\langle A_y \rangle = 0$, then
the transformed $A'_y = - \left(\frac{1}{e}\right)\frac{\alpha}{L}$ again 
satisfies
$F'_{\mu\nu}=0$.
For the Minkowski space $M^5$, therefore, the non-vanishing zero-mode is 
just of pure gauge, and a theory with $\langle A_y \rangle \neq 0$ is
gauge-equivalent to the case of $\langle A_y \rangle = 0$.

The above argument has to be modified, once $M^5$ is replaced by $M^4\times
 S^1$.  What is crucial here is the property that the compact space $S^1$
 is a nonsimply-conneted space.
To see how a contant $A_y$ can be physical, let us write $A_y$ as
\begin{eqnarray}
A_y(x_\mu,y) = A_y^c + A_y^q(x_{\mu},y),
\end{eqnarray}
where $A_y^c$ denotes the constant background field (zero-mode) or $A_y^c = 
\langle A_y\rangle$, while
$A_y^q$ denotes quantum fluctuation around the background.
Solving the Dirac equation in the presence of $A_y^c$, we get a spinor
$\psi$ with non-trivial b.c.:
\begin{eqnarray}
\psi(x_\mu,y + L) = \psi(x_\mu,y) e^{ie\oint A_y^c dy} , 
\end{eqnarray}
even if we assume that $\psi$ satisfies periodic b.c. in the absence of
$A_y^c$.  The Wilson-loop along the $S^1$, $e\oint A_y^c dy = e A_y^c L$, (corresponding to 
$\alpha$) is a gauge invariant quantity, and may be understood as $e\Phi$
with $\Phi$ being the ``magnetic flux'' inside the $S^1$ (we 
do not have to worry about what the ``inside'' of $S^1$ means).
This is a sort of Aharonov-Bohm effect.  Since $e\Phi$ has gauge
invariant physical meaning as the magnetic flux , the theory should be
different if the value of $A_y^c$ is different.

In particular, the vacuum energy will depend on $A_y^c$, which in turn
means that $A_y^c$ and equivalently $\alpha$ can be fixed dynamically as
the value which realizes the ground state.  Namely we can calculate the
vacuum energy density $V(A_y^c)$as a function of $A_y^c$ (``gauge
potential'') and the vacuum state is realized as the minimum
point of $V(A_y^c)$.  $V(A_y^c)$ is obtained by evaluating $\mbox{Tr ln}
(i\gamma^M D_M)$, where the covariant derivative is due to $A_y^c$,
$D_y = \partial_y - ie A_y^c$,
just as in the case of the calculation of the Coleman-Weinberg potential
in the background field method. The result is
\begin{eqnarray}
V(A_y^c)&=&- \frac{L^{-D}}{2^{\frac{D-1}{2}}\pi^{\frac{D}{2}}\Gamma 
(\frac{D}{2})} \int_0^\infty \! ds \ s^{D-1} \nonumber \\ 
        & &\times\ln \left[1 - 2 \cos(e A_y^c L) e^{-\sqrt{s^2+(mL)^2}} +
e^{-2\sqrt{s^2+(mL)^2}}\right]. 
\end{eqnarray}
The
divergent contribution for $L \rightarrow \infty$, which is independent
of $A_y^c$ and has no physical significance, has been subtracted in Eq.(15).
%
%
\begin{figure}
\begin{center}
\includegraphics{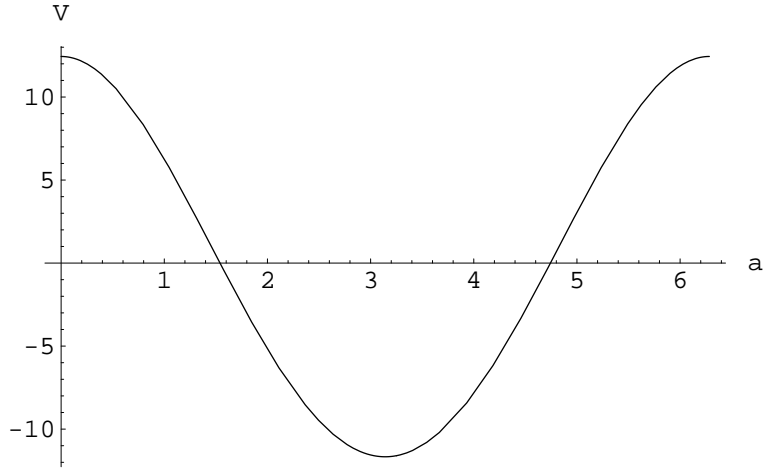}
\caption{
The vacuum energy density $V$ in the unit of $\frac{1}{2\sqrt{2} \pi^2 L^4}$ 
as a function of $a$, with $a$ standing for $\alpha = e A_y^c L$.}
\end{center}
\end{figure}
$V(A_y^c)$ is plotted as a function of $\alpha = e A_y^c L$ (for $D = 4$
and $m = 0$) in Fig.3 . In this approach $\mass^2$ is given as the
coefficient of the term quadratic in $A_y^c$ :
\begin{eqnarray}
\mass^2 = \frac{d^2 V}{dA_{y}^{c2}} \ .
\end{eqnarray}
When $ e \displaystyle A_y^c L$ is identified with $\alpha$, this 
definition of $\mass^2$ coincides with the
direct calculation of the 2-point function, Eq.(7). As is shown in Fig.3, 
the value $\alpha
\simeq 0.46\pi$ corresponds to the point where $\frac{\displaystyle d^2
V}{\displaystyle dA_{y}^{c2}} = 0$, but it does not correspond to the 
stationary point with $\frac{\displaystyle d V}{\displaystyle dA_{y}^{c}} = 0$.
The minimum of $V(A_y^c)$ is at $\alpha = \pi$, where 
$\mass^2 > 0$. Thus, unfortunately $\mass^2 =0$ is not achieved 
dynamically, in the toy model $M^4\times
 S^1$.

Calculations of the vacuum energy under constant background gauge 
fields
have previously been made by Hosotani in $M^3 \times S^1$ with the fermion 
mass $m=0$ \cite{Hosotani}.
His purpose was to study the gauge symmetry breaking 
due to the VEV of non-Abelian gauge fields.  After this pioneering work many papers have appeared to discuss the Hosotani mechanism in various models \cite{Hetrick}. 
Our purpose was not to discuss the spontaneous gauge symmetry breaking 
($\langle A_y \rangle$ does not break U(1) ) but to discuss $\mass$. 
     
\vspace{0.5 cm}
\leftline{\bf 4. The Higgs mass on $S^2$}
\vspace{0.2 cm}

A natural question to be raised is why non-vanishing gauge boson mass
$m_{A_y}^2 = \mass^2 \neq 0$ became possible, without contradicting the
local gauge invariance. One may naively argue that if $m_{A_y}^2$ is present the
mass term is not invarant under a gauge transformation, $A_y \rightarrow
A_y +c$ ($c$ is a constant). We, however, have shown that $A_y$ and $A_y+c$
are no longer gauge equivalent to each other, unless $c=(2\pi/eL)n$
($n$ is an integer), since $A_y$ and $A_y+c$ correspond to different
values of magnetic flux in the sence of the A-B effect. This is
why the vacuum energy depends on $A_y^c$, generating the ``curvature''
$\frac{\displaystyle d^2 V}{\displaystyle dA_{y}^{c2}} = m_{A_y}^2$.

Thus it may be reasonable to expect that the local gauge invariance
does work to guarantee $\mass^2=0$, provided the compactified space is a
simply-connected manifold, not allowing the penetration of magnetic
flux.  We will illustrate this by calculating $\mass^2$ in the
case where the compactified space is $S^2$. For 
simplicity, we ignore 4-dimensional space-time and calculate
$\mass^2$ just on $S^2$, taking scalar QED theory as an example. We
assume that the matter scalar is massless.

According to the method we took above, we calculate the vacuum energy
density under the background photon field $A_m$.  This
generates the effective action for $A_m$, i.e. $I = \int d^{2}y \sqrt{g}
\L_{eff} = \mbox{Tr}\ln\{(D_m+ieA_m)(D^m+ieA^m)\}$ and $D^m$ is
covariant under the general coordinate transformation ($g_{mn}$: the
metric of $S^2$).  The $\mbox{Tr}\ln$ is obtained by path-integration
over the scalar field under the background $A_m$. On the other hand,
gauge and general coordinate invariance imply that the term in
$\L_{eff}$, which are quadratic in $A_m$, should be written in a form
\begin{equation}
\L_{eff} = a(\partial_m A_n - \partial_n A_m)^2 + bg_{mn}A^m A^n,
\end{equation}
where the kinetic term takes the ordinary form so that it respects local 
gauge symmetry,
as the coefficient $a$
is a dimensionlesss parameter and should not be affected by whether
$R$(the radius of $S^2$) is $\infty$ or finite. To obtain the
mass-squared $b$, which should be identified with $\frac{1}{2}\mass^2$,  we 
take a specific choice of background, $A_\theta =0$
and $A_\phi=\mbox{constant}$ ( $(\theta,\phi)$ are angular coordinates on
$S^2$). Then $a$ term disappears and
\begin{equation}
b = \frac{1}{2}\mass^2 = \mbox{Tr}\{-e^2g_{\phi\phi}(\Delta_{S_2})^{-1}
               + 2e^2\partial^2_{\phi}(\Delta_{S^2})^{-2}   \}
    / (\int d^2y \sqrt{g}g_{\phi\phi}),
\end{equation}
where the numerator comes from the second derivative of the $\mbox{Tr} \ln$ 
with respect to
$A_{\phi}$, and $\Delta_{S^2} \equiv g^{mn} D_m \partial_n$ is the
laplacian on $S^2$.
The first term of the numerator can be evaluated by taking the trace
with respect to configuration space coordinates as
\begin{equation}
-e^2 \left(\int d^2y \sqrt{g}g_{\phi\phi}\right) \cdot G(0,0)
\end{equation}
 in terms of the propagator $G(y,y')$ ($\Delta_{S^2} G(y,y') =
\delta(y-y')$),
\begin{equation}
G(y,y')= \sum_{l=0}^{\infty}\sum_{m=-l}^{l}
 \left\{ -\frac{1}{l(l+1)} \right\} Y_l^m (\theta^{\prime}, 
\phi^{\prime})^{\ast}Y_l^m (\theta, \phi).
\end{equation}
The second term is simply given as the sum of eigenvalues of
$\partial_\phi^2(\Delta_{S^2})^{-2}$, i.e. \\ 
$- {m^2 a^4}/{[l(l+1)]^2}$.
  We thus get
\begin{equation}
\mass^2 = 2e^2 \sum_{l,m} \frac{1}{l(l+1)} |Y_l^m(0,0)|^2 -4e^2a^4 
\left\{\sum_{l,m} \frac{m^2}{[l(l+1)]^2}
    / \int d^2y \sqrt{g}g_{\phi\phi}\right\}.
\end{equation}
Since $\displaystyle \sum_{m}|Y_l^m(0,0)|^2 = \frac{2l+1}{4\pi}$,
 $\displaystyle \sum_{m} m^2 = \frac{1}{3} l(l+1)(2l+1)$ and  
 $\displaystyle \int d^2y \sqrt{g}g_{\phi\phi} = \frac{8\pi}{3}a^4$, we
 finally get
\begin{equation}
\mass^2 = \frac{e^2}{2\pi} \sum_{l} \frac{2l+1}{l(l+1)}
   - \frac{e^2}{2\pi}\sum_l \frac{2l+1}{l(l+1)} = 0 .
\end{equation}
To be consistent with our naive expectation, the Higgs mass-squared
$\mass^2$ identically vanishes irrespective of the size of the compactified space $S^2$.
It needs some further study to see whether the vanishing $\mass$ is still realized 
when 4-dimensional
space-time is added to $S^2$.

\vspace{0.5 cm}
\leftline{\bf 5. Concluding remarks}
\vspace{0.2 cm}

We studied the possibility to solve the
gauge hierarchy problem thanks to the local gauge symmetry 
in the framework of higher-dimensional gauge
theories. We first took a toy model, i.e., QED in $M^D \times S^1$
space-time.  Because of the
local gauge symmetry, at the tree level the Higgs mass-squared $\mass^2$ 
vanishes automatically and the ultraviolet quadratic divergence in quantum
correction to $\mass^2$ also cancells out when non-zero (massive) modes are
all summed up in the internal loop. 
We, however, found a finite correction to $\mass^2$. 
When periodic boundary condition (b.c.) is taken for the fermion, small $\mass$ of the weak scale was argued to be realized either by a relatively large compactification 
(length) scale, 
say $R \sim 1/(1TeV)$ or so, or by a heavy fermion of mass $m > 1/R$. 
The $\mass^2$ was shown to be very sensitive to the b.c. and 
to vanish for a specific choice of the b.c.. 

We also discussed that the b.c. may be regarded as the consequence of 
a sort of Aharonov-Bohm effect, and can be fixed dynamically.
 Unfortunately, it turned out that such fixed b.c. does not 
lead to $\mass^2 = 0$. 
Finally we investigated the case where the compactified space is
 $S^2$. The A-B effect, which is known to play an
essential role to yield the finite $\mass^2$, is not allowed for
the simply-connected space $S^2$, and our calculation shows that $\mass^2$ 
identically vanishes.

The analysis given in this paper are all in toy models. In order to make
our mechanism viable, further discussions are obviously necessary to
clarify e.g. whether the mechanism also works in realistic GUT models
and how a Higgs field belonging to fundamental representation is 
realized starting from the adjoint repr. of gauge field.

\noindent {\bf Acknowledgment}

We would like to thank K. Kikkawa, 
T. Kubota for stimulating discussions and the members of high energy theory group 
at Kobe university for useful conversations.  This work has been supported by the 
Grant-in-Aid for Scientific Research (09640361) from the Ministry of Education, 
Science and Calture, Japan.


\begin{thebibliography}{99}
\bibitem{Sakai}
S. Dimopoulos and H. Georgi,
{\sl Nucl. Phys.},
{\bf B193}(1981) 150;
N. Sakai,
{\sl Z. Phys.},
{\bf C11}(1981) 153.

\bibitem{techni}
L. Susskind,
{\sl Phys. Rev.}
{\bf D20} (1979) 2619;
S. Weinberg,
{\sl Phys. Rev.}
{\bf D19} (1979) 1279;
V.A. Miranskii, M. Tanabashi and K. Yamawaki,
{\sl Phys. Lett.}
{\bf 221B} (1989) 177.

\bibitem{Inoue}
K. Inoue, A. Kakuto, and H. Takano,
{\sl Progr. Theor. Phys.}
{\bf 75} (1986) 664;
R. Barbieri, G. Dvali, and A. Strumia,
{\sl Nucl. Phys.}
{\bf B391} (1993) 487.

\bibitem{'tHooft}
G. 't Hooft,
{\sl The proceedings of Cargese Summer Inst.}
(1979) 135.

\bibitem{SUSY}
For the review of supersymmtery see, for example, \\ 
J. Wess and J. Bagger, {\sl Supersymmetry and 
Supergravity}, Princeton Univ. Press, 1983; 
H.P. Nills, {\sl Phys. Rep.} {\bf C110} (1984) 2.

\bibitem{Tom}
D.J. Toms,
{\sl Ann. Phys.}
{\bf 129} (1980) 334.

\bibitem{Ford}
L.H. Ford,
{\sl Phys. Rev.}
{\bf D21} (1980) 933.

\bibitem{Inami}
P. Candelas and S. Weinberg, {\sl Nucl. Phys.} {\bf B237} (1984) 397; T. Inami and K. Yamagishi,
{\sl Phys. Lett.}
{\bf 143B} (1984) 115; 
K. Kikkawa, T. Kubota, S. Sawada and M. Yamasaki, 
{\sl Nucl. Phys.}
{\bf B260} (1985) 429. \\ 
For the review of Kaluza-Klein type theories see, 
for example, \cite{KK}.  

\bibitem{KK}
{\sl Modern Kaluza-Klein Theories} ed. by 
T. Appelquist, A. Chodos and P.G.O. Freund, 
Addison-Wesley (1987).  

\bibitem{Hosotani}
Y. Hosotani,
{\sl Phys. Lett. }
{\bf 126B} (1983) 309;
{\sl Phys. Lett.}
{\bf 129B} (1983) 193.

\bibitem{Antoniadis}
I. Antoniadis,
{\sl Phys. Lett.}
{\bf 246B} (1990) 377; 
I. Antoniadis, C. Mu$\tilde{n}$oz and M. Quir$\acute{o}$s,
{\sl Nucl. Phys.}
{\bf B397} (1993) 515; 
I. Antoniadis, K. Benakli and M. Quir$\acute{o}$s,
{\sl Phys. Lett.}
{\bf 331B} (1994) 313.

\bibitem{Large}
N. Arkani-Hamed, S. Dimopoulos and G. Dvali, 
{\sl hep-ph/9803315}; {\sl hep-ph/9807344}; 
K.R. Dienes, E. Dudas and T. Gherghetta, 
{\sl hep-ph/9803466}; {\sl hep-ph/9806292}; 
I. Antoniadis, N. Arkani-Hamed, S. Dimopoulos and G. Dvali, 
{\sl hep-ph/9804398}; 
G. Shiu and S.-H. H. Tye, {\sl hep-th/9805157}; 
C. Bachas, {\sl hep-ph/9807415}.

\bibitem{Hetrick}
J.E. Hetrick and Y. Hosotani,
{\sl Phys. Rev. }
{\bf D38} (1988) 2621; 
{\sl Phys. Lett. }
{\bf 230B} (1989) 88; 
A. Higuchi and L. Parker, 
{\sl Phys. Rev. }
{\bf D40} (1988) 2853; 
J.E. Hetrick and C.-L. Ho,
{\sl Phys. Rev. }
{\bf D40} (1989) 4085.

\end{thebibliography}
\end{document}